**Category specificity of N170 response recovery speeds for faces and Chinese characters**

Xiaoli Ma, Cuiyin Zhu, Chenglin Li, Xiaohua Cao

Department of Psychology, Zhejiang Normal University, Jinhua, China. 321001

Correspondence to: Xiaohua Cao, Department of Psychology, Zhejiang Normal University. Jinhua, China. 321001
Email: xiaohua.cao@vanderbilt.edu    Tel: +86 057982282549    Fax: +86 057982282549

**Abstract**

Neural selectivity of N170 responses is an important phenomenon in perceptual processing; however, the recovery times of neural selective responses remain unclear. In the present study, we used an adaptation paradigm to test the recovery speeds of N170 responses to faces and Chinese characters. The results showed that recovery of N170 responses elicited by faces occurred between 1400 and 1800 ms after stimuli onset, whereas recovery of N170 responses elicited by Chinese characters occurred between 600 and 800 ms after stimuli onset. These results demonstrate category-specific recovery speeds of N170 responses involved in the processing of faces and Chinese characters.

**Introduction**

Recent research has used neural adaptation to characterize neuronal populations involved in visual processing (Andrews & Ewbank, 2004; Feuerriegel, Churches, & Keage, 2015; Grill-Spector & Malach, 2001; Koutstaal et al., 2001).. Neural adaptation refers to the reduction of neural activity in response to repeated presentation of physically or categorically identical stimuli (Grill-Spector, Henson, & Martin, 2006). This adaptation effect is a powerful tool in neuroscience research due to its stimulus-specific nature (see Grill-Spector et al., 2006, for a review). In a commonly used adaptation paradigm, an adaptor stimulus and a subsequent test stimulus are presented during each trial. Recent studies have employed this adaptation paradigm to examine category-sensitive processing measured by the N170 response (Eimer, Gosling, Nicholas, & Kiss, 2011; Eimer, Kiss, & Nicholas, 2010), known as the N170 adaptation effect. In the N170 adaptation paradigm, the adaptor and test stimulus are presented successively with a specified inter-stimulus interval (ISI). The N170 adaptation effect is measured by comparing N170 amplitudes evoked by the same test stimulus when preceded by different categories of adaptor stimuli.

The N170 adaptation paradigm has been used to characterize neural responses to objects of expertise (e.g. faces, words). For example, many studies have demonstrated that N170 amplitudes are smaller for face tests preceded by face adaptors than by non-face adaptors (e.g., houses), suggesting a face-specific N170 adaptation effect (Amihai, Deouell, & Bentin, 2011; Cao, Jiang, Gaspar, & Li, 2014a; Kovacs, Zimmer, Volberg, Lavric, & Rossion, 2013; Nemrodov & Itier, 2011). However, many studies have failed to find N170 adaptation effects as a result of face stimulus repetition (e.g.

Eimer, 2000a; Engst, Martń-Loeches, & Sommer, 2006; Kuehl, Brandt, Hahn, Dettling, & Neuhaus, 2013; Neumann & Schweinberger, 2008; Schweinberger, Pickering, Jentzsch, Burton, & Kaufmann, 2002). A closer look at contradicting studies of face-specific N170 adaptation reveals that ISIs are typically longer, between 1200 and 2000 ms, in negative reports (Eimer, 2000a; Neumann & Schweinberger, 2008) than in studies confirming N170 adaptation effects (200 ms; Eimer et al., 2011; Kovacs et al., 2006). This pattern suggests that the N170 neural response elicited by the adaptor stimulus does not recover during a short ISI and can therefore reduce the N170 response to the test stimulus, resulting in an adaptation effect. However, the N170 response to the adaptor recovers completely during a long ISI and thus cannot affect the response to the test stimulus. Hence, recovery of N170 responses may explain the lack of adaptation effects in studies using long ISIs.

Importantly, Kuehl et al. (2013) investigated this issue by varying the ISIs from 400 to 2000 ms and found that the N170 response recovers almost completely during an ISI of 1600 ms. This result suggests that the ISI is an important factor in the N170 adaptation effect. It also suggests that the N170 adaptation paradigm with varying ISIs represents a sensitive method with which to investigate the recovery speed of neural selectivity responses.

Similar to face-specific N170 adaptation effects, word-related N170 adaptation effects have been found in studies using short ISIs (e.g., 200) (Cao, Ma, & Qi, 2015; Cao et al., 2014a; Cao, Jiang, Li, & He, 2014b)), but not in studies using long ISIs (e.g., Maurer et al., 2008, jittered between 1250 and 1750 ms; Mercure et al.,

2011,1000 ms). These results also suggest that N170 neural responses elicited by words do not recover during a short ISI and can thus reduce the response to a test stimulus word, but show complete recovery during a long ISI. However, the recovery time of the word-related N170 response remains unclear.

Taken together, these results suggest that the N170 adaptation effect is a stable phenomenon in the processing of different visual expert stimuli when a short ISI is used, whereas the N170 response recovers almost completely within 2000 ms from stimuli onset. However, these studies raise two questions regarding the recovery speed of N170 responses elicited by objects of expertise. Firstly, the time necessary for complete recovery of N170 responses elicited by non-face expert stimuli (e.g., words) has not been determined. Secondly, it is unclear whether different categories of stimuli (e.g., faces or words) have similar N170 response recovery speeds. In the present study, we addressed these questions using an adaptation paradigm to test the recovery speeds of N170 responses to the two most common objects of expertise (faces and words).

**Methods**

*Participants*

A total of 40 subjects (22 male; age 20.5 ± 2.05 years) were recruited from a local university and paid for their participation. All subjects were right-handed with normal or corrected-to-normal vision. Written consent was obtained from each subject, and the study design was approved by the ethics committee of Zhejiang Normal University. None of the participants were familiar with the experimental design of the study.

*Stimuli*

Stimuli consisted of grayscale pictures of faces, Chinese characters, and line drawings. Face stimuli consisted of images of 60 individuals (30 male and 30 female) displaying a neutral facial expression, which were selected from a standard set of faces used in our laboratory. External features (hair and ears) were replaced with an oval contour using Adobe Photoshop CS6 (San Jose, CA). Sixty high-frequency Chinese characters with a left-right configuration and 8–14 strokes were chosen from the Modern Chinese Frequency Dictionary (1986) and presented in a common font (Song). Control adaptor stimuli consisted of 60 grayscale images of line drawings selected from Shu, Cheng & Zhang (1989). Face stimuli were $198 \times 218$ pixels, subtending angles of $4.7° \times 5.5°$ from a viewing distance of 80 cm. Chinese characters and line drawing stimuli were $198 \times 198$ pixels, subtending angles of $4.7° \times 5.0°$ from a viewing distance of 80 cm.

*Procedure*

Participants sat in a dimly lit room at a distance of 80 cm from the 17" CRT monitor ($1024 \times 768$ pixel resolution) on which stimuli were presented against a dark gray background. E-Prime 2.0 software was used for stimuli presentation and behavioral response recording (Psychology Software Tools, Pittsburgh, PA).

In each trial, an adaptor stimulus and a test stimulus were presented sequentially for 200 ms each, with ISIs of 200, 400, 600, 800, 1200, or 1600 ms. The inter-trial intervals varied randomly between 2500 and 2700 ms. Faces (F) or Chinese characters (C) were used as test stimuli. Each test stimulus was preceded by an adaptor stimulus of the same category (face-face [FF] or Chinese character-Chinese character [CC] trials), or the control category (line drawing-face [LF] or line drawing-Chinese character [LC] trials). Half of the participants participated in face adaptation testing (FF and LF trials) and the remaining participants participated in Chinese character adaptation testing (CC and LC trials). Each participant was tested using 12 conditions, consisting of same-category or control category trials, each with 6 different ISIs (see above). Participants performed 12 testing blocks, each with one of 6 different ISIs. Each block consisting of 33 same-category trials and 33 control category trials with the same ISI, presented randomly. The 12 blocks were presented in random order. Of the 66 trials in each block, 60 were non-target trials, in which no response was required. The remaining 6 trials per block were target trials, in which the stimulus shape was outlined in red. Target trials were used in order to maintain participants' attention to the task. The red-outlined target stimulus was presented with equal probability as the adaptor stimulus or test stimulus in the target trials. The target trials were randomly intermixed with the non-target trials. Participants were instructed to press a response button following presentation of the second stimulus when they detected a target stimulus.

*Electroencephalograpy (EEG) and data analysis*

EEG was performed using a 128-channel HydroCel Geodesic Sensor Net (Electrical Geodesics, Inc., Eugene, OR), with an electrode placed on the vertex (Cz) serving as a reference for online recording. Electrode impedances were maintained below 50 kΩ. Signals were digitized at a 500 Hz sampling rate and amplified with a 0.1–200 Hz elliptical bandpass filter. EEG data were digitally filtered offline with a 0.3–30 Hz bandpass filter and sorted into epochs from 200 ms before to 800 ms after stimulus onset with a baseline from 50 ms before until 50 ms after stimulus onset. Trials with artifacts exceeding ± 100 μV were rejected. Trials containing eye movement artifacts were also excluded from event-related potential (ERP) averaging. A minimum of 45 good trials for each stimulus category was required to retain a participant in the analyses. Data from two female participants in face adaptation testing and one male and one female participant in Chinese character adaptation testing were excluded from future analyses. The remaining EEG data were re-referenced to the average of all channels.

EEG data were analyzed from non-target trials only. Groups of channels over the left occipitotemporal (O1, 65, T5; channel 65 between O1 and T5) and right occipitotemporal regions (O2, 90, T6; channel 90 between O2 and T6), where the N170 components were maximal, were analyzed (Cao et al., 2014). In order to reduce the number of levels in the statistical analyses, peak amplitudes and latencies were then averaged across the three channels for each hemisphere. EEG waveforms were averaged separately for each presentation condition. Based on visual inspection of individual data, the N170 time window was defined as 150–210 ms after adaptor stimuli onset and 140–220 ms after test stimuli onset. N170 responses to adaptor stimuli were analyzed using a repeated measures multivariate analysis of variance (MANOVA) for trial type (same-category or control category trials), adaptation condition (face or Chinese character adaptation), and hemisphere (left, right). N170 responses to test stimuli were analyzed using a MANOVA for trial type, ISI, adaptation condition, and hemisphere. To clarify the adaptation effects with each ISI for both faces and Chinese characters, we analyzed N170 amplitudes elicited by test stimuli separately for each ISI.

**Results**

*Behavioral results*

No main effects of accuracy rate and response time (RT) were found. The mean accuracy rates in face adaption trials and Chinese character adaptation trials were 96% and 97%, respectively. The RTs in face adaption trials and Chinese adaptation trials were 496 ms and 454 ms, respectively.

*N170 response amplitudes for adaptor stimuli*

We found a main effect of trial type ($F(1,34) = 111.56$, $p < 0.001$, $\eta_p^2 = 0.766$), such that N170 responses elicited by objects of expertise (faces or Chinese characters) were larger than those elicited by non-expert objects (line drawings) (Figure 1).

*N170 response latencies for adaptor stimuli*

A significant interaction was found between trial type and adaptation condition ($F(1,34) = 39.79$, $p < 0.001$, $\eta_p^2 = 0.539$). Further analysis revealed that N170 latencies for face stimuli were longer than latencies for line drawings ($F(1,17) = 23.32$, $p < 0.001$, $\eta_p^2 = 0.578$). Additionally, N170 latencies for Chinese characters were shorter than latencies for line drawings ($F(1,17) = 17.44$, $p = 0.001$, $\eta_p^2 = 0.506$) (Figure 1).

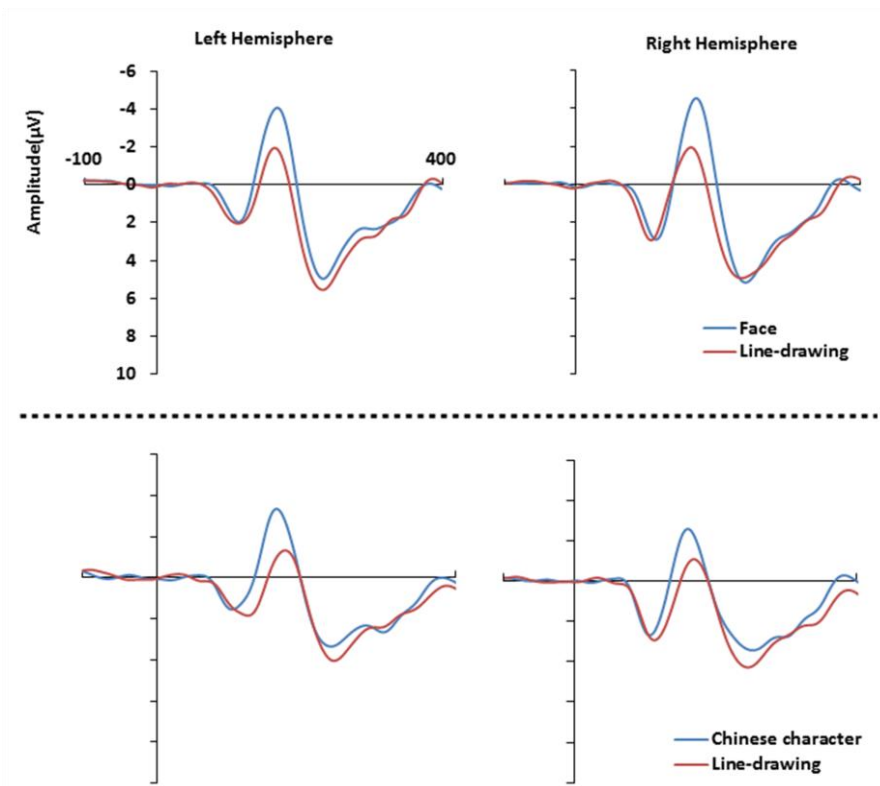

Figure 1. Averaged N170 waveforms recorded from the left and right hemispheres elicited by adaptor stimuli during face adaptation testing (top row) and Chinese character adaptation testing (bottom row)

*N170 response amplitudes for test stimuli*

We found main effects of trial type ($F(1,34) = 48.30$, $p < 0.001$, $\eta_p^2 = 0.587$) and ISI ($F(5,170) = 18.94$, $p < 0.001$, $\eta_p^2 = 0.358$). A significant interaction was found between trial type and adaptation condition ($F(1,34) = 16.09$, $p < 0.001$, $\eta_p^2 = 0.321$).

A significant interaction was also found between trial type and ISI ($F(5,170) = 12.93$, $p < 0.001$, $\eta_p^2 = 0.276$). A significant interaction was also found between adaptation condition and ISI ($F(5,170) = 2.74$, $p = 0.036$, $\eta_p^2 = 0.074$). Finally, a three-way interaction was found among trial type, adaptation condition, and hemisphere ($F(1,34) = 6.65$, $p = 0.014$, $\eta_p^2 = 0.164$).

*N170 response latencies for test stimuli*

We found significant main effects of trial type ($F(1,34) = 67.93$, $p < 0.001$, $\eta_p^2 = 0.666$) and ISI ($F(5,170) = 73.28$, $p < 0.001$, $\eta_p^2 = 0.683$). A three-way interaction was found among trial type, ISI, and adaptation condition ($F(5,170) = 8.84$, $p < 0.001$, $\eta_p^2 = 0.206$).

Further analysis revealed that N170 response latencies for face stimuli were shorter in same-category trials than in control category trials ($F(1,17) = 32.07$, $p < 0.001$, $\eta_p^2 = 0.654$). Additionally, a main effect of ISI for face adaptation was found ($F(5,85) = 44.02$, $p < 0.001$, $\eta_p^2 = 0.721$). Post-hoc *t* tests revealed that N170 latencies with an ISI of 200 ms were significantly shorter than N170 latencies with longer ISIs (all $p < 0.001$), and N170 latencies with an ISI of 400 ms were significantly shorter than N170 latencies with ISIs of 600, 800, and 1200 ms (all $p < 0.05$) (Table 1; Figures 2 and 4).

Chinese character adaptation testing revealed main effects of trial type ($F(1,17) = 37.21$, $p < 0.001$, $\eta_p^2 = 0.686$) and ISI ($F(5,85) = 33.98$, $p < 0.001$, $\eta_p^2 = 0.667$). A significant interaction was found between trial type and ISI ($F(5,85) = 8.71$, $p < 0.001$,

$\eta_p^2 = 0.339$). A post-hoc $t$ test revealed that N170 latencies were shorter in same-category trials than in control category trials when the ISI was 200 ms ($t(17) = 5.20$, $p < 0.001$), 400 ms ($t(17) = 3.66$, $p = 0.002$), or 1600 ms ($t(17) = 3.18$, $p = 0.005$) (Table 1; Figures 3 and 4).

In order to clearly identify adaptation effects for both face and Chinese character stimuli with each ISI, we analyzed the data for each ISI separately. The results are shown in Table 1 and Figures 2-4.

*200 ms ISI*

We found a main effect of trial type ($F(1,34) = 48.22$, $p < 0.001$, $\eta_p^2 = 0.586$). A significant interaction between trial type and adaptation condition was found ($F(1,34) = 10.60$, $p = 0.003$, $\eta_p^2 = 0.238$). A three-way interaction among trial type, hemisphere, and adaptation condition was also found ($F(1,34) = 4.24$, $p = 0.047$, $\eta_p^2 = 0.111$). Further analysis revealed that N170 responses to faces as test stimuli were larger when preceded by line drawing adaptors than by face adaptors ($F(1,17) = 33.28$, $p < 0.001$, $\eta_p^2 = 0.662$). Similarly, N170 responses to Chinese characters as test stimuli were larger following line drawing adaptors than responses following Chinese character adaptors ($F(1,17) = 15.57$, $p = 0.001$, $\eta_p^2 = 0.478$).

Table 1 Amplitude and Latency values of the N170 in Response to test stimulus for the two Adaptation Conditions in the ISIs

|  |  |  | 200 ms | | 400 ms | | 600 ms | | 800 ms | | 1200 ms | | 1600 ms | |
|---|---|---|---|---|---|---|---|---|---|---|---|---|---|---|
|  |  |  | SC | DC | SC | DC | SC | DC | SC | DC | SC | DC | SC | DC |
| FA | L | Amplitudes (μV) | -3.97 (2.85) | -6.26 (4.02) | -3.18 (2.58) | -4.55 (2.95) | -2.44 (2.41) | -3.45 (3.10) | -3.38 (2.59) | -4.42 (3.08) | -3.52 (2.71) | -4.12 (3.01) | -4.06 (3.08) | -4.18 (3.29) |
|  |  | Latencies (ms) | 187 (19.31) | 185 (13.91) | 171 (12.27) | 175 (15.10) | 167 (12.57) | 172 (13.34) | 165 (12.65) | 169 (13.30) | 166 (13.56) | 164 (8.13) | 170 (15.25) | 172 (12.62) |
|  | R | Amplitudes (μV) | -4.21 (2.54) | -7.00 (3.32) | -2.88 (2.37) | -4.65 (2.81) | -2.76 (2.31) | -3.68 (2.69) | -3.41 (2.75) | -4.02 (3.32) | -3.67 (3.30) | -4.48 (3.19) | -3.90 (2.93) | -4.40 (3.12) |
|  |  | Latencies (ms) | 186 (12.42) | 189 (11.67) | 172 (10.54) | 174 (10.09) | 169 (10.60) | 172 (10.77) | 168 (11.47) | 171 (10.47) | 168 (11.65) | 173 (11.23) | 172 (11.70) | 174 (11.83) |
| CA | L | Amplitudes (μV) | -4.28 (2.73) | -5.56 (3.39) | -2.79 (2.48) | -4.24 (2.63) | -2.89 (2.74) | -2.95 (3.00) | -3.24 (2.79) | -3.12 (2.88) | -3.21 (2.84) | -3.49 (3.03) | -3.50 (3.08) | -3.58 (3.03) |
|  |  | Latencies (ms) | 176 (14.49) | 190 (12.82) | 168 (11.49) | 173 (13.27) | 163 (13.39) | 163 (9.63) | 165 (14.61) | 162 (11.34) | 162 (14.33) | 172 (13.10) | 162 (9.21) | 165 (10.88) |
|  | R | Amplitudes (μV) | -4.20 (2.71) | -4.76 (3.08) | -2.21 (2.18) | -2.98 (2.77) | -2.09 (2.49) | -2.12 (2.92) | -2.29 (2.72) | -1.83 (2.53) | -2.63 (2.93) | -2.39 (3.13) | -2.79 (3.20) | -2.75 (2.85) |
|  |  | Latencies (ms) | 177 (11.78) | 193 (14.97) | 165 (12.84) | 172 (13.13) | 162 (15.33) | 162 (10.35) | 159 (11.23) | 161 (10.30) | 160 (10.97) | 162 (11.40) | 164 (11.59) | 171 (17.50) |

Note "SC" and "DC" were respectively the abbreviations of "Same Condition and" and Different Condition; "FA" and "CA" were respectively the abbreviations of "Face Adaptation" and Chinese character Adaptation. "L" and "R" were respectively refer to "Left Hemisphere and" Right Hemisphere.

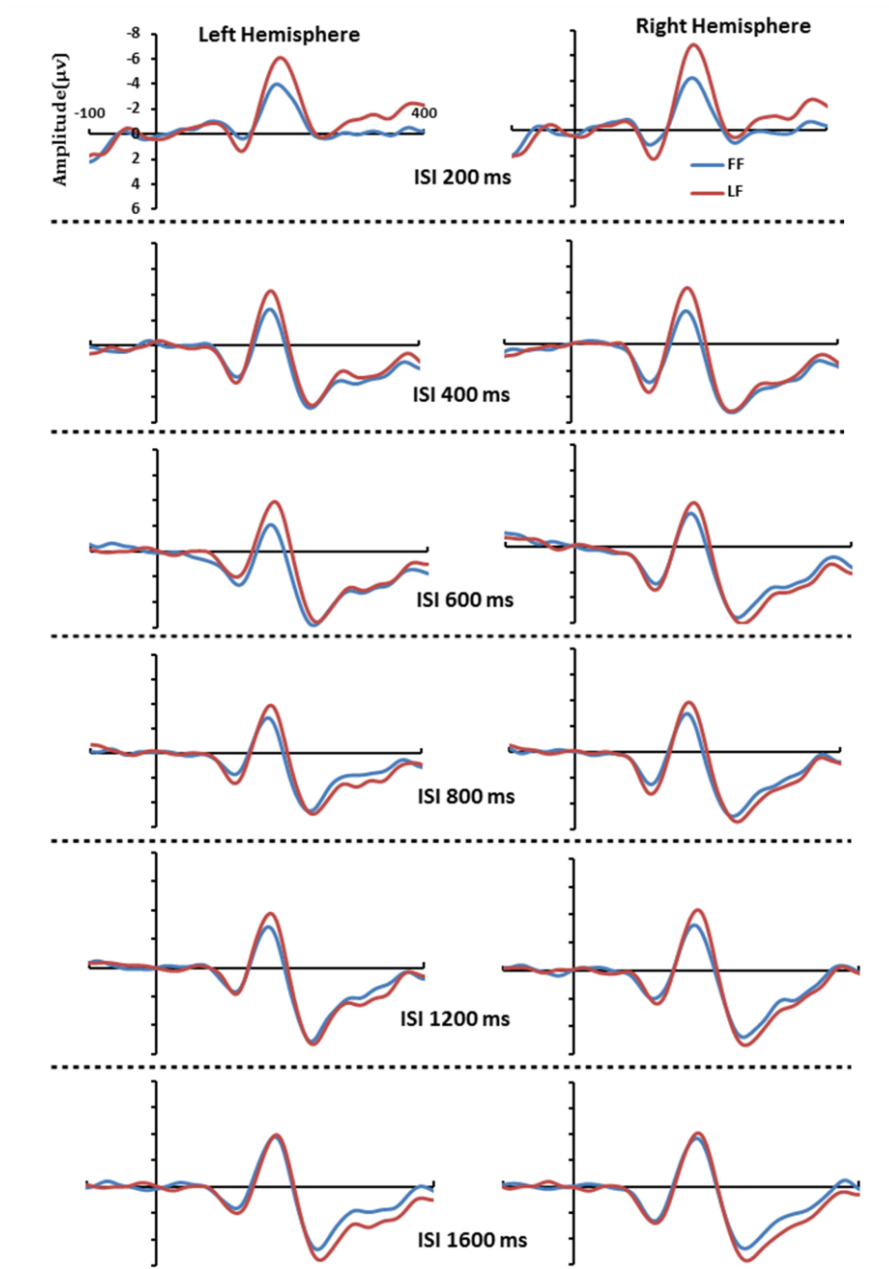

Figure 2. Averaged N170 waveforms elicited by test stimuli with different ISIs during face adaptation testing.

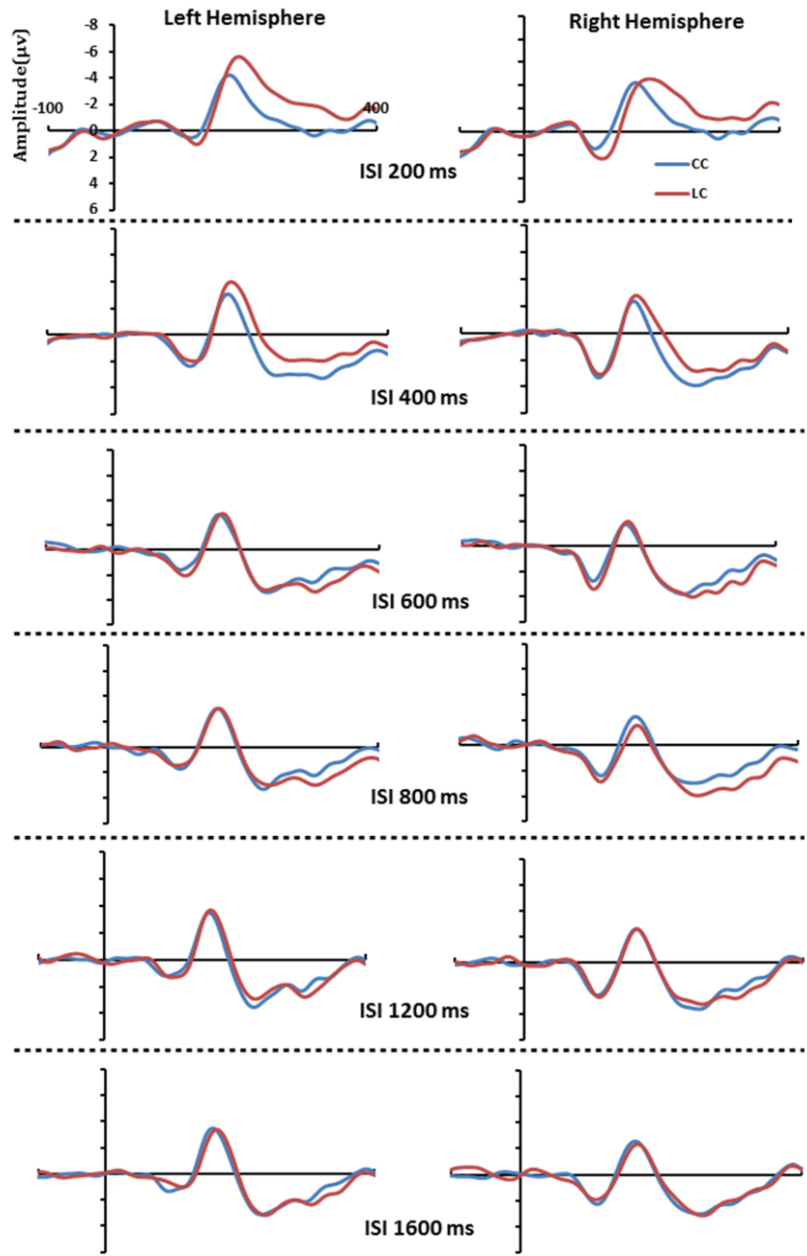

Figure 3. Averaged N170 waveforms elicited by test stimuli with different ISIs during Chinese character adaptation testing.

*400 ms ISI*

There was a main effect of trial type ($F(1,34) = 62.69$, $p < 0.001$, $\eta_p^2 = 0.648$). A three-way interaction among trial type, hemisphere, and adaptation condition was also found ($F(1,34) = 5.19$, $p = 0.029$, $\eta_p^2 = 0.132$). Further analysis revealed that N170 responses to faces as test stimuli were larger when preceded by line drawing adaptors than by face adaptors ($F(1,17) = 47.57$, $p < 0.001$, $\eta_p^2 = 0.737$). Similarly, N170 responses to Chinese characters as test stimuli were larger when the adaptor was a line drawing than when the adaptor was a Chinese character ($F(1,17) = 19.61$, $p < 0.001$, $\eta_p^2 = 0.535$). Furthermore, N170 responses to Chinese characters were larger in the left hemisphere than in the right hemisphere ($F(1,17) = 9.04$, $p = 0.008$, $\eta_p^2 = 0.347$).

*600 ms ISI*

We found a main effect of trial type ($F(1,34) = 62.69$, $p < 0.001$, $\eta_p^2 = 0.648$). A significant interaction between trial type and adaptation condition was also found ($F(1,34) = 4.83$, $p = 0.035$, $\eta_p^2 = 0.124$). A post-hoc *t* test revealed that N170 responses elicited by faces in control category trials were larger than in same-category trials ($t(17) = 3.64$, $p = 0.002$). However, N170 responses to Chinese characters as test stimuli were similar in same-category and control category trials ($t(17) = 0.30$, $p = 0.765$).

*800 ms ISI*

We found a significant interaction between trial type and adaptation condition ($F(1,34) = 9.60$, $p = 0.004$, $\eta_p^2 = 0.220$). A post-hoc *t* test revealed that N170 responses elicited by faces were larger in control category trials than in same-category trials ($t(17) = 2.76$, $p = 0.014$). However, N170 responses to Chinese characters as test stimuli were similar in same-category and control category trials ($t(17) = 1.46$, $p = 0.163$).

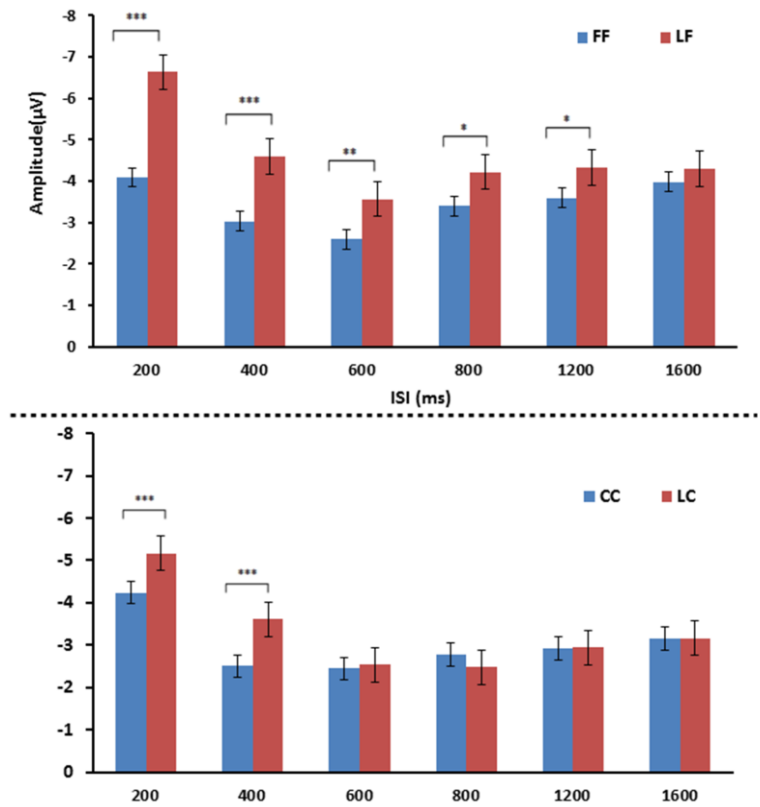

Figure 4. Average N170 amplitudes elicited by test stimuli with different ISIs during face adaptation testing (top row) and Chinese character adaptation testing (bottom row). * $p < 0.05$, ** $p < 0.01$, *** $p < 0.001$.

*1200 ms ISI*

There was a main effect of trial type ($F(1,34) = 4.22$, $p = 0.048$, $\eta_p^2 = 0.110$). A marginally significant interaction between trial type and adaptation condition was also found ($F(1,34) = 3.79$, $p = 0.060$, $\eta_p^2 = 0.100$). A post-hoc *t* test revealed that N170 responses elicited by face stimuli were larger in control category trials than in same-category trials ($t(17) = 2.45$, $p = 0.025$). However, N170 responses to Chinese characters as test stimuli were similar in same-category and control category trials ($t(17) = 0.09$, $p = 0.929$).

*1600 ms ISI*

No main effects or interactions were found with an ISI of 1600 ms.

**Discussion**

Our results demonstrated that N170 responses elicited by both face and Chinese character stimuli were stronger than those elicited by line drawings, consistent with previous studies (Bentin, Allison, Puce, Perez, & McCarthy, 1996; Cao, Li, Zhao, Lin, & Weng, 2011;Cao, Jiang, Li, & He, 2014; Cao, Jiang, Li, Xia, & Floyd, 2015; Maurer, Brem, Bucher, & Brandeis, 2005; Rossion & Jacques, 2008; Rossion, Joyce, Cottrell, & Tarr, 2003). We also showed that N170 responses to faces as test stimuli were larger in trials with control adaptors than in trials with face adaptors when the ISIs were 200, 400, 600, 800, or 1200 ms. This result replicates face-specific N170 adaptation effects found in previous studies (Cao, Jiang, Gaspar, et al., 2014; Cao et al., 2015; Eimer et al., 2011; Eimer et al., 2010; Fu et al., 2012; Jacques, d'Arripe, & Rossion, 2007; Kovacs et al., 2006; Kuehl et al., 2013; Nemrodov & Itier, 2011; Nemrodov & Itier, 2012). Importantly, N170 response amplitudes with faces as test stimuli were similar between trials using control or face adaptors when the ISI was 1600 ms, suggesting that face-evoked N170 responses completely recovered between 1400 and 1800 ms after stimulus onset. Kuehl et al. (2014) also found that N170 responses showed nearly complete recovery during an ISI of 1600 ms. Taken together, these results suggest that the recovery speed of face-specific N170 responses is similar across cultures.

Importantly, we also demonstrated that N170 responses to Chinese characters as test stimuli were larger in trials with control adaptors than in trials with Chinese character adaptors when the ISI was 200 or 400 ms. These results are consistent with N170 adaptation effects for Chinese characters that have been found in recent studies (Cao et al., 2014a, 2014b; Cao et al., 2015). Furthermore, our results showed that N170 amplitudes triggered by Chinese characters as test stimuli were similar between trials using control or Chinese character adaptors when the ISI was 600, 800, 1200, or 1600 ms, suggesting that N170 responses elicited by Chinese characters completely recover between 600 ms and 800 ms after Chinese characters onset. These results help to explain previous contradictory findings regarding word-specific N170 response adaptation. Our data obtained from trials with a short ISI (200 or 400 ms) are consistent with many recent studies showing that word stimuli can induce N170 adaptation with a 200-ms ISI (Cao, Jiang, Gaspar, et al., 2014; Cao, Jiang, Li, et al., 2014; Cao et al., 2015). However, Our results of trials using an ISI longer than 400 ms are consistent with the results obtained by Fu, Feng, Guo, Luo, & Parasuraman (2012), Maurer et al. (2008) and Mercure et al. (2011), which failed to find word-specific N170 adaptation effects with ISIs longer than 1000ms. Interestingly, Feng et al. (2013) and Fu et al. (2012) found that the N170 adaptation to Chinese characters occurred with the ISI 500-800 ms random, and our results showed that the N170 response recover with ISI longer than 600ms. Together with our results and the results from the two studies noted above, it implied that the trials with ISI 500-600 ms in their studies may mainly contributed to the N170 adaptationin.

By systematically varying the ISI, we show that face-evoked N170 responses can completely recover between 1400 and 1800 ms from stimulus onset. Our results also demonstrate for the first time that word-evoked N170 responses can completely recover between 600 and 800 ms after stimulus onset. Thus, our results show that the recovery speed of the N170 response is category-specific, with the response to Chinese characters recovering more quickly than the response to faces.

These results raise the question of why recovery of the N170 response to faces is slower than that of the response to Chinese characters. The most likely explanation is that N170 responses elicited by faces have larger amplitudes than those elicited by words and therefore require more time to return to baseline levels, as demonstrated by recent studies. For example, Maurer et al. (2008) found that responses to face stimuli were stronger than responses to word stimuli in both blocked and alternating testing conditions. Mercure et al. (2008) also found larger N170 responses to faces than to words when the stimuli were large and had high resolution. Mercure et al. (2011) demonstrated that N170 responses to faces were larger than those to words when preceded by a different category of stimuli. Aranda et al. (2010) found that N170 responses to faces were larger than those to Spanish words. Similar results have been obtained using non-alphabetic languages such as Chinese, showing that N170 responses to faces are larger than those to Chinese characters under normal display conditions. For instance, Feng et al. (2013) and Fu et al. (2012) demonstrated that N170 responses to faces were larger than those to Chinese characters with a long ISI. Our recent results using a similar adaptation paradigm also showed that N170 responses to face adaptor stimuli were larger than those to Chinese character adaptor stimuli. Taken together, the above studies support the hypothesis that large N170 responses to face stimuli require more time to return to baseline levels. Future studies should be designed to directly test this hypothesis.

However, the reasons for the difference in absolute N170 amplitudes across objects of expertise (e.g., faces and words) remain unclear. The face-elicited N170 is thought to reflect structural encoding of face stimuli (Bentin & Deouell, 2000; Eimer, 2000b). Since all faces present the same first-order configuration (two eyes above a nose above a mouth; Maurer, Le Grand, & Mondloch, 2002); the resulting similarity between different faces may be one reason for the strong N170 response to faces. Previous literature has shown that word-evoked N170 responses may represent an index of visual word form recognition (Maurer, Brandeis, & McCandliss, 2005). In comparison to faces, different words share a less common configuration, which may explain why N170 responses elicited by words are smaller than those elicited by faces. Future studies should directly test this hypothesis.

In conclusion, our results demonstrate that the N170 response to a face stimulus completely recovers between 1400 ms and 1800 ms after stimulus onset. Moreover, our results provide the first demonstration that the N170 response to a Chinese character stimulus can recover completely between 600 ms and 800 ms after stimulus onset. Importantly, this demonstration of category-specific N170 recovery speeds extends the view that the N170 response indexes different cognitive mechanisms for the processing of face and orthographic stimuli. Our results also emphasize the importance of using an appropriate ISI when investigating N170 adaptation effects for different object categories.

## Acknowledgments

This study was supported by Social Science Foundation of China (14BYY064) and National Science Foundation of China (31571159).